\documentclass{cpcauth}
\usepackage{graphicx}

\begin{document}

\begin{frontmatter}

\title{Pseudo-First-Order Transition in Interacting Self-avoiding Walks and Trails}

\author{Thomas Prellberg}
\ead{thomas.prellberg@tu-clausthal.de}
\address{Institut f\"ur Theoretische Physik, Technische Universit\"at Clausthal,
Arnold-Sommerfeld Stra\ss e 6, D-38678 Clausthal-Zellerfeld, Germany}
\author{Aleksander L. Owczarek}
\ead{aleks@ms.unimelb.edu.au}
\address{Department of Mathematics and Statistics, University of Melbourne, Parkville, Vic 3052, Australia}

\begin{abstract}
The coil-globule transition of an isolated polymer has been well
established to be a second-order phase transition described by a
standard tricritical O(0) field theory. We present 
Monte-Carlo simulations of interacting self-avoiding walks and 
interacting self-avoiding trails
in four dimensions which provide compelling evidence
that the approach to this (tri)critical point is
dominated by the build-up of first-order-like singularities
masking the second-order nature of the coil-globule transition. 
\end{abstract}

\end{frontmatter}

Much work has been done on polymer collapse models in the physically important
dimensionalities two and three. In addition to general tri-critical scaling theory,
results from conformal field theory and exactly solvable models have given a thorough
understanding of the polymer collapse transition, leaving but a few open questions. At the 
upper critical dimension $d_u=3$ results from field-theoretical work and simulations also
confirm the tri-critical scaling behaviour \cite{grassberger1995a-a,grassberger1997a-a,hager1999a-a}.

Until recently \cite{owczarek1998a-:a}, polymer collapse above the upper critical dimension has attracted little attention,
presumably because it was generally accepted that it is described by standard mean field theory and
therefore should be of little interest. However, our work in \cite{prellberg2000a-:a,owczarek2000a-:a,prellberg2001a-:a}
uncovers a surprisingly interesting scenario.

Mean-field theory is generally applicable to second-order phase
transitions above their upper critical dimension, and so is believed
to provide an adequate description of the approach to such critical
points. One type of transition where mean-field theory should hold are
tri-critical points
\cite{lawrie1984a-a} for dimension $d>3$. The region around a
tri-critical point in general dimension
is described by crossover scaling forms, where
quantities depending on two relevant parameters can be essentially
described by functions of a single scaling combination of those two
parameters. 

The application of the mean-field theory of a tricritical point to
polymer collapse predicts that at the transition point the polymer
actually behaves as if it were a random walk.  In the thermodynamic
limit, one expects a weak transition with a jump in the
specific heat $\alpha=0$. For finite polymer
length there is no sharp transition for an isolated polymer
and so this mean-field transition is rounded and shifted. 
In four and higher dimensions one may expect pure mean-field behaviour 
with a crossover exponent of $1/2$ \cite{gennes1975a-a}.

On the other hand, for $d>3$ Sokal \cite{sokal1994a-a} has pointed out
that the alternative method of analysing collapse which has been shown
to be equivalent to the field theoretic approach, namely the continuum
Edwards model, has difficulties: in fact, an analysis of the Edwards
model shows that the crossover exponent is given by $\phi_E=2-d/2$,
which for $d=4$ gives $\phi_E=0$! In passing we note here that the same
analysis predicts the shift of the $\theta$-point, defined say via the
universal ratio of the radius of gyration to the end-to-end distance
equalling its Gaussian value, should scale as $N^{-(d/2 -1)}$ so
$\psi_E=(d/2 -1)\neq\phi_E$. This difference between the shift and the
crossover exponent implies that strict crossover scaling has broken
down. Of course, the theoretical fact that the swollen phase should
also be Gaussian for $d>4$ does raise the suspicion that the analysis
of the Edwards model for polymer collapse may be subtle for $d>3$.

\begin{figure}
\begin{center}
\includegraphics[width=7cm]{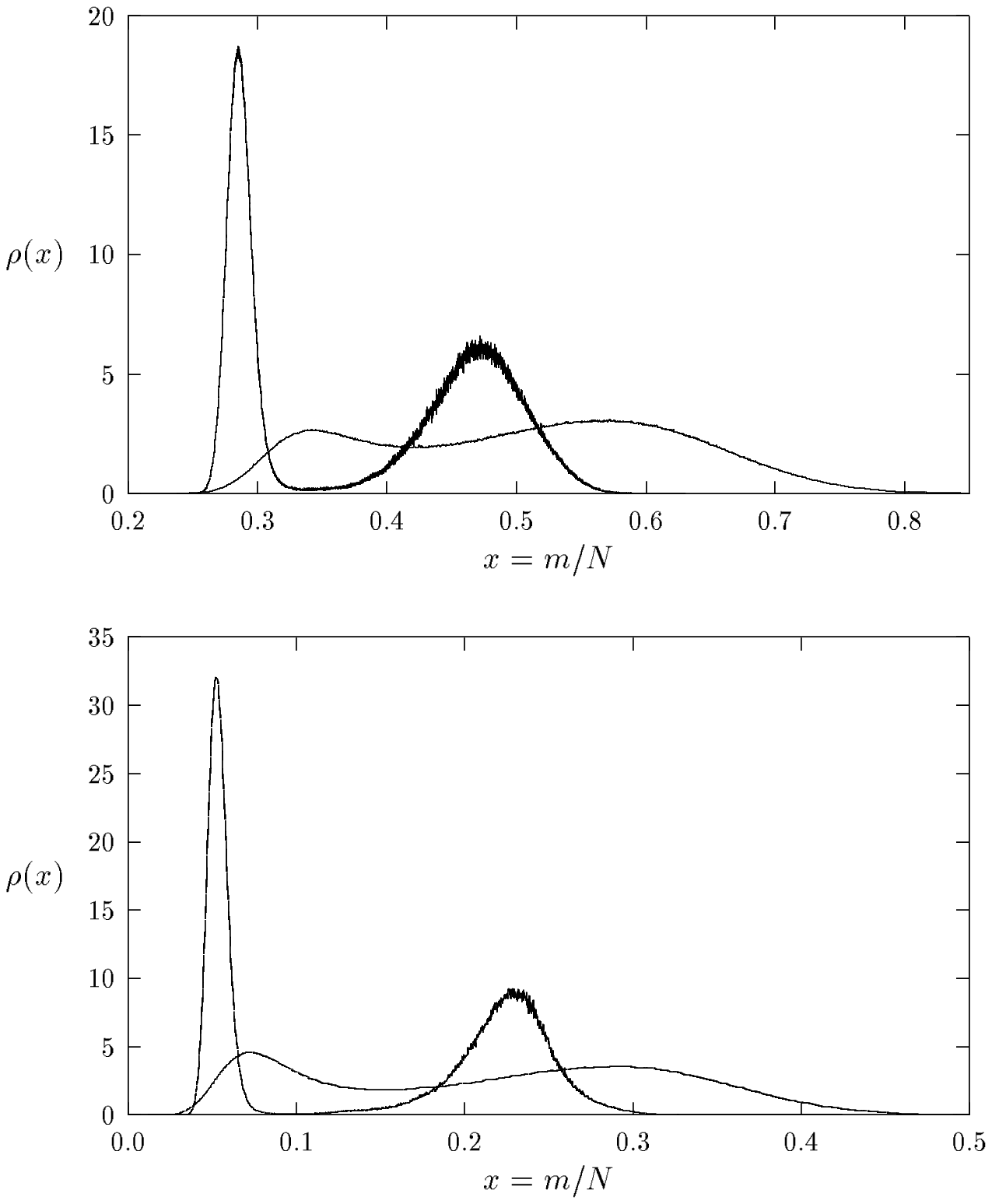}
\caption{\it
Internal energy density distributions for interacting self-avoiding walks at lengths 2048 and 16384 (above) and interacting self-avoiding trails at lengths 512 and 4096 (below) 
on the four-dimensional hyper-cubic lattice, at their respective transition temperatures. 
The more highly peaked distribution is associated with the longer respective length 
(Figures 11 from \cite{prellberg2000a-:a} and 7 from \cite{prellberg2001a-:a}).}
\label{fig14} 
\end{center}
\end{figure}

\begin{figure}
\begin{center}
\includegraphics[width=7cm]{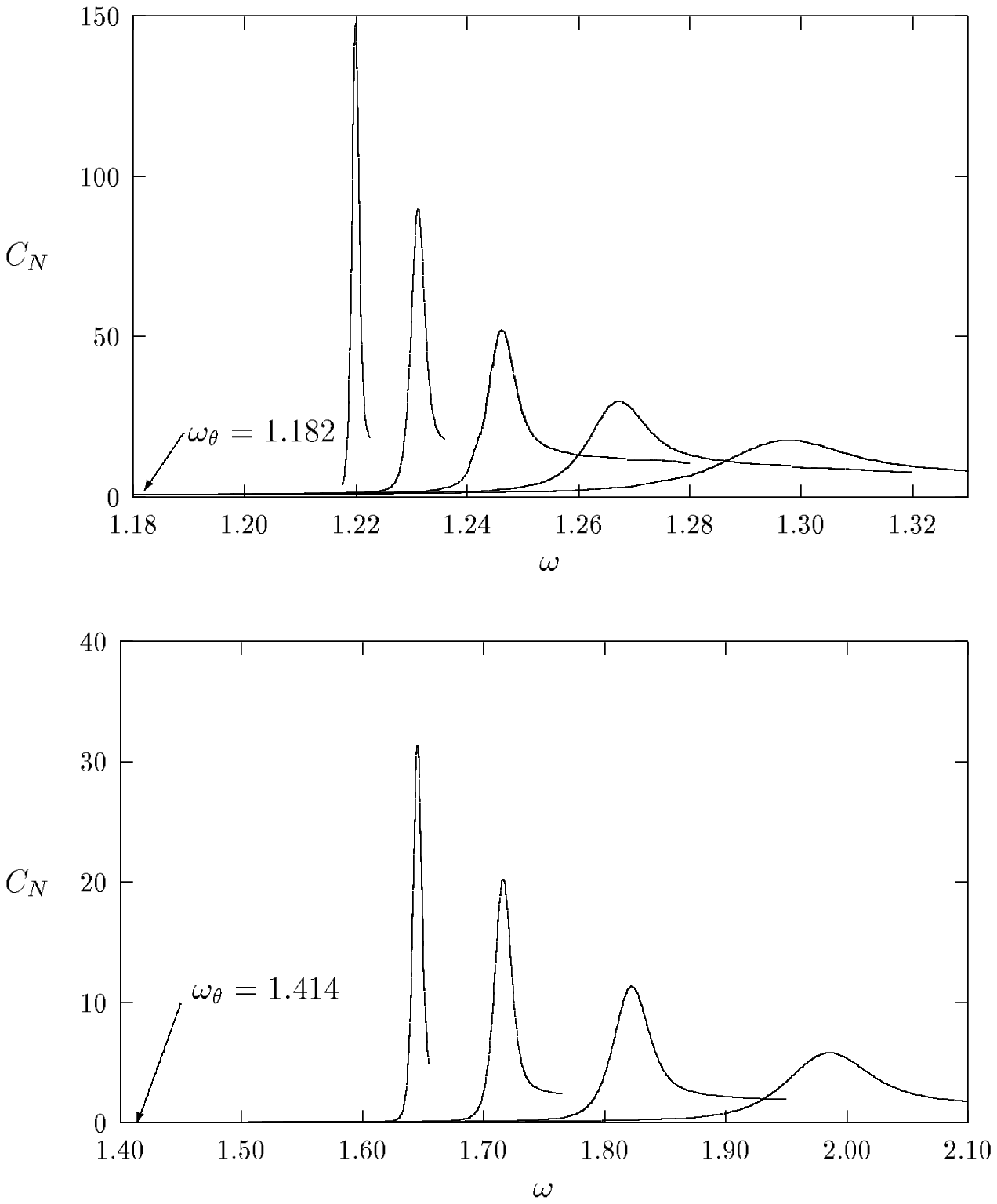}
\caption{\it
Specific heat $C_N$ versus $\omega$ for interacting self-avoiding walks at lengths 1024, 2048 , 4096, 8192 and 16384 (above) 
and for interacting self-avoiding trails at lengths 512, 1024, 248, and 4096 (below) on the four-dimensional hyper-cubic lattice. 
(Figures 9 from \cite{prellberg2000a-:a} and 5 from \cite{prellberg2001a-:a}).}
\label{fig15} 
\end{center}
\end{figure}

To consider such issues, we have simulated two lattice models of polymer collapse. The first is the canonical model of
interacting self-avoiding walk (ISAW), where one associates an attractive interaction with non-consecutive nearest-neighbour
interactions of a self-avoiding walk. (For obvious reasons it is not sufficient to model the collapse by simply weakening 
the self-avoidance, as this allows for the possibility of an unphysical buildup of density in small spacial regions.)

The model of interacting self-avoiding trails (ISAT) is yet another plausible lattice model of polymer collapse,
with self-avoidance restricted to bonds and attractive interaction incorporated via contacts. There is some evidence that
while self-avoiding trails are in the same universality class as self-avoiding walks the corresponding interacting models
may have different scaling at their collapse points. For instance, simulations on the square lattice show that there are 
logarithmic corrections to scaling at the ISAT $\theta$-point \cite{owczarek1995a-:a}.

Using PERM, a clever generalisation of a kinetic growth algorithm \cite{grassberger1997a-a}, we have simulated
interacting self-avoiding walks and interacting self-avoiding trails on the four-dimensional
hyper-cubic lattice \cite{prellberg2000a-:a,prellberg2001a-:a}, and report on
the implications for collapse scaling in \cite{owczarek2000a-:a}. PERM builds upon the
Rosenbluth-Rosenbluth method \cite{rosenbluth1955a-a}, in which configurations
are generated by simply growing an existing configuration kinetically but
overcomes the exponential ``attrition'' and re-weighting needed in
this approach by a combination of enrichment and pruning strategies. It turns
out that PERM is highly efficient for simulations of polymers near the $\Theta$-point.

We find that there is a rather dramatic breakdown of the simple crossover scaling for 
the case of the coil-globule transition of an isolated polymer.
It is likely that the build up of the tri-critical point is
through the forming of singularities that have more in common with a
(non-critical) first-order transition! However, this can be explained
by a different kind of mean-field approach (not starting with an
explicitly tri-critical Landau functional); moreover, the region
around the tri-critical point needs to be described by more complex
scaling forms. This second issue is in fact separate from the
first-order nature of the scaling approach: we speculate that this
behaviour is intimately related to the general description of systems
where mean-field theory is used, so may have more general
applicability.

The main evidence for this scenario stems from the internal energy density distribution 
near the collapse transition, which is shown in Figure \ref{fig14}. 
The character of that transition is particularly intriguing;  we find a
distinct double peak distribution for the internal energy, which
becomes \emph{more} pronounced as the chain
length is increased.  This would seem to suggest a first-order
transition. If this were the case there would be a delta function peak
forming in the specific heat but we find that while a peak is indeed
forming it does not seem to be growing linearly with the size of the
polymer, see Figure \ref{fig15}. Moreover, there is a $\theta$-point scaling region
distinct from the collapse transition (the location of which is indicated in 
Figure \ref{fig15}), a scenario which is incompatible with a first-order transition.

Fortunately there is a (suitably extendable) theoretical framework
that is consistent with the evidence we
present. This framework was explained in a paper by Khokhlov
\cite{khokhlov1981a-a} who applied the mean-field approach of
Lifshitz, Grosberg and Khokhlov 
\cite{lifshitz1968a-a,lifshitz1976a-a,lifshitz1978a-a} to arbitrary
dimensions. This theory is based on a phenomenological free energy in which 
the competition between a
bulk free energy of a dense globule and its surface tension drive the
transition. Until recently \cite{owczarek1993c-:a} the consequences of
this surface free energy were largely ignored in the polymer
literature. 

The implications of this theory for polymer collapse above the upper
critical dimension are described in \cite{prellberg2000a-:a,owczarek2000a-:a}.
The major conclusion is that the finite-size
character of the coil-globule transition in four dimensions is
first-order despite the thermodynamic limit being probably adequately
described by mean-field tricritical behaviour. We propose to call this 
a {\em pseudo-first-order transition}. One consequence is the breakdown
of conventional tri-critical scaling; the single-variable scaling
form needs to be replaced by more complex scaling forms.

When comparing the data for interacting trails and walks in more detail,
we note further that the bimodal distribution
emerges for trails at much shorter configurations than for walks, so that the peaks in
the distribution for trails at length $N=512$ are already more
pronounced than the peaks in the distribution for walks at length
$N=2048$.  To quantify this observation, we turn to the scaling
predictions of LGK theory. An important parameter in the theory is the
quotient $a^d/v$, where $a$ is the mean-square distance between two
subsequent monomers (repeated unit element of the polymer: equivalent
to occupied sites of the lattice model) along a chain and $v$ is the
effective excluded volume of a monomer, defined via the vanishing of
the second virial coefficient at the $\theta$-temperature. For instance, 
the shift of the transition temperature
is given by
\begin{equation}
{\omega_{c,N}-\omega_\theta\over\omega_\theta}\sim\left({\tilde sa^4\over Nv}\right)^{1/3}
\end{equation}
where $\tilde s$ is a constant proportional to the quotient of the third 
virial coeffficient and the excluded volume squared.
We estimate that
$N^{1/3}(\omega_{c,N}-\omega_\theta)$ asymptotes to $3.4(1)$ for trails,
and for walks we estimate for the same quantity the value $0.92(3)$.
Identifying $a$ with the lattice constant, 
which in both models is set equal to one, we can get a rough estimate for 
the relative size of the effective excluded volume $v$ in both models. 
We obtain
\begin{equation}
{v_{SAT}\over v_{SAW}}\approx 0.03 {\tilde s_{SAT}\over\tilde s_{SAW}}
\end{equation}
and thereby quantify the intuitive notion that the excluded volume
effect is numerically ``weaker'' in trails than in walks, though of
the same basic type.

\end{document}